\documentclass[11pt,epsf,letterpaper]{article}%
\usepackage{subfig}
\usepackage{color}
\usepackage{amsmath}
\usepackage{amsfonts}
\usepackage{verbatim}
\usepackage{amssymb}
\usepackage{graphicx}
\usepackage{epstopdf}
\usepackage{mathrsfs}
\usepackage{epsfig}
\usepackage{cite}%
\providecommand{\U}[1]{\protect\rule{.1in}{.1in}}
\providecommand{\U}[1]{\protect\rule{.1in}{.1in}}

\begin{document}

\title{\flushright{\small UAI-PHY-16/10} \\\center{{\huge {Hairy AdS Solitons}}}}
\author{Andr\'{e}s Anabal\'{o}n$^{(1)}$, Dumitru Astefanesei$^{(2)}$ and David
Choque$^{(3)}$\\
\\\textit{$^{(1)}$Departamento de Ciencias, Facultad de Artes Liberales and} \\\textit{ Facultad de Ingenier\'{\i}a y Ciencias, Universidad Adolfo
Ib\'{a}\~{n}ez, } \\\textit{Av. Padre Hurtado 750, Vi\~{n}a del Mar, Chile.} \\
\\\textit{$^{(2)}$Instituto de F\'\i sica, Pontificia Universidad Cat\'olica de
Valpara\'\i so,} \\\textit{Casilla 4059, Valpara\'{\i}so, Chile.} \\
\\\textit{$^{(3)}$Universidad T\'{e}cnica Federico Santa Mar\'{\i}a, Av.
Espa\~{n}a 1680, Valpara\'{\i}so, Chile.}\\[1mm]
\texttt{{\small andres.anabalon@uai.cl,dumitru.astefanesei@pucv.cl,brst1010123@gmail.com}
}}
\maketitle

\begin{abstract}
We construct exact hairy AdS soliton solutions in Einstein-dilaton gravity
theory. We examine their thermodynamic properties and discuss the role of these solutions for the existence of first order phase transitions for hairy black holes. The negative energy density associated to hairy AdS solitons can be interpreted  as 
the Casimir energy that is generated in the dual filed theory when the fermions are antiperiodic on the compact coordinate.
\end{abstract}

\newpage


\section{Introduction}

We construct analytic neutral hairy soliton solutions in Anti de Sitter (AdS)
spacetime and discus their properties. This analysis is important in the
context of AdS/CFT duality \cite{Maldacena:1997re} because bulk solutions
correspond to `phases' of the dual field theory \cite{Witten:1998zw}.

There is by now a huge literature on (locally) asymptotically AdS solutions in
both phenomenological models and consistent embedding in supergravity. We will
consider theories of gravity coupled to a scalar field with potential
$V(\phi)$. AdS spacetime is not globally hyperbolic, which means that the
evolution is well defined if the boundary
conditions are imposed. In particular, since for the same self-interaction there exist
many boundary conditions for the scalar field (that may or may not break the
conformal symmetry in the boundary), one can `design' a specific field theory
\cite{Hertog:2004ns} with a given effective potential \cite{Hertog:2004ns,
Hertog:2004rz, Anabalon:2015vda}.

Different foliations of AdS spacetime lead to different definitions of time
and so to distinct Hamiltonians of the dual field theory. Since the classical
(super)gravity background, with possible $\alpha^{\prime}$ corrections, is
equivalent to the full quantum gauge theory on the corresponding slice, one
expects physically inequivalent dual theories for different foliations.
Indeed, when the horizon topology of the black hole is Ricci flat and there
are no compact directions, there are no first order phase transitions similar
to the Hawking-Page \cite{Hawking} phase transitions that exist for the
spherically symmetric black holes. However, when
some of the spatial directions are compactified on a circle asymptotically,
one expects the existence of a negative Casimir energy of the
non-supersymmetric field theory that `lives' on the corresponding topology.
Horowitz and Myers have shown in \cite{Horowitz:1998ha} that, indeed, there
exists a (bulk) gravity solution dubbed `AdS soliton' with a lower energy than
AdS itself. This solution was obtained by a double analytic continuation (in
time and one of the compactified angular directions) of the planar black hole. This
fits very nicely with the proposal of Witten \cite{Witten:1998zw} that a
non-supersymmetric Yang-Mills gauge theory can be described within AdS/CFT
duality by compactifying one direction and imposing anti-periodic boundary
conditions for the fermions around the circle.

Hairy neutral AdS solitons were previously analysed (see, e.g.
\cite{Brihaye:2013tra, Ogawa:2011fw, Shi:2016bxz, Kleihaus:2013tba,
Brihaye:2012ww, Smolic:2015txa, Cadoni:2011yj}), though most of these studies
are using numerical methods. Hence, it would be interesting to find examples
of analytic hairy AdS solitons and investigate their generic properties. In
recent years, analytic regular neutral hairy black holes in AdS were
constructed, e.g. \cite{Acena:2013jya, Acena:2012mr, Feng:2013tza,
Wen:2015xea, Lu:2013ura, Fan:2015tua, Fan:2015ykb} and so one expects that
constructing analytic soliton solutions could be also possible. We use some
particular exact planar hairy black hole solutions in four and five dimensions
of \cite{Acena:2013jya,Acena:2012mr} and obtain the corresponding solitons by
using a double analytic continuation as in \cite{Horowitz:1998ha}. The hairy
AdS solitons are the ground state candidates of the theory
\cite{Woolgar:2016axs}.

Since the AdS soliton is the solution with the minimum energy within these
boundary conditions \cite{Galloway:2001uv,Galloway:2002ai}, it is natural to
investigate the existence of phase transitions with respect to this thermal
background. In the nice work \cite{Surya:2001vj}, it was shown that there
exist first order phase transitions between planar black holes and the AdS
soliton. We construct the hairy AdS soliton and compute their mass by using
the counterterm method of Balasubramanian and Kraus
\cite{Balasubramanian:1999re} supplemented with extra counterterms for the
scalar field as was proposed in \cite{Anabalon:2015xvl}. We investigate then
the existence of first order phase transitions with respect to the hairy AdS
soliton and discuss the effect of `hair' on the thermodynamical behaviour.

\section{Hairy AdS soliton}

In this section we construct exact hairy AdS soliton solutions in four and
five dimensions and compute their energy. In five dimensions
\cite{Acena:2013jya,Acena:2012mr}, we obtain a new hairy black hole solution,
which corresponds to a parameter $\nu$ that at first sight makes the moduli
potential divergent. However, by taking the right limit, we show that the
theory is in fact well defined and the solution is regular.


\subsection{AdS soliton}

We start with a short review of \cite{Surya:2001vj}, though, to connect this
analysis with the rest of the paper, the computations are done by using the
counterterm method of Balasubramanian and Kraus \cite{Balasubramanian:1999re}.

We consider the usual AdS gravity action supplemented with the gravitational
counterterm proposed in \cite{Balasubramanian:1999re}
\begin{equation}
I[g_{\mu\nu}]=\int_{\mathcal{M}}d^{4}x\left(  R-2\Lambda\right)  \sqrt
{-g}+2\int_{\partial\mathcal{M}}{d^{3}x~K\sqrt{-h}}-\int_{\partial\mathcal{M}%
}{d^{3}x~\frac{4}{l}\sqrt{-h}} \label{actionSchw}%
\end{equation}
where $\Lambda=-3/l^{2}$ is the cosmological constant ($l$ is the radius of
AdS), $16\pi G_{N}=1$ with $G_{N}$ the Newton gravitational constant, the
second term is the Gibbons-Hawking boundary term, and the last term is the
gravitational counterterm. Here, $h$ is the determinant of the induced
boundary metric and $K$ is the trace of the extrinsic curvature. The planar
black hole solution is
\begin{equation}
ds^{2}=-\biggl{(}-\frac{\mu_{b}}{r}+\frac{r^{2}}{l^{2}}\biggr{)}dt^{2}%
+\biggl{(}-\frac{\mu_{b}}{r}+\frac{r^{2}}{l^{2}}\biggr{)}^{-1}dr^{2}%
+\frac{r^{2}}{l^{2}}(dx_{1}^{2}+dx_{2}^{2}) \label{bh}%
\end{equation}
where $\mu_{b}$ is the mass parameter and we consider the compactified
coordinates $0\leq x_{1}\leq L_{b}$ and $0\leq x_{2}\leq L$. The normalization
is such that the time coordinate and the coordinates $x_{1}$ and $x_{2}$ have
the same dimension and so the analytic continuation for obtaining the AdS
soliton produces the same boundary geometry.

The role of the counterterm is to cancel the infrared divergence of the action
so that the final result is finite:
\begin{equation}
I^{E}_{b}=\frac{2L L_{b}\beta_{b}}{l^{4}}\biggl{(}-r^{3}_{b}+\frac{\mu
_{b}l^{2}}{2}\biggr{)}=-\frac{L L_{b}\beta_{b}r^{3}_{b}}{l^{4}}
\label{Schwact}%
\end{equation}
The horizon radius is denoted by $r_{b}$ and $\beta_{b}$ is the periodicity of
the Euclidean time that is related to the temperature of the black hole by:
\begin{equation}
T = \beta_{b}^{-1}=\frac{(-g_{tt})^{^{\prime}}}{4\pi}\biggr{\vert}_{r=r_{b}%
}=\frac{3r_{b}}{4\pi l^{2}} \label{schtemp}%
\end{equation}
Using the usual thermodynamic relations and free energy $F = I^{E}_{b}/
\beta_{b}$, we obtain the energy and entropy of the planar black hole:
\begin{equation}
\label{enT}E=-T^{2}\frac{\partial I_{b}^{E}}{\partial T}=\frac{2LL_{b}\mu_{b}%
}{l^{2}}%
\end{equation}
\begin{equation}
S=-\frac{\partial(I_{b}^{E}T)}{\partial T}=\frac{LL_{b}r_{b}^{2}}{4l^{2}G_{N}%
}=\frac{\mathcal{A}}{4G_{N}}%
\end{equation}

The AdS soliton solution was obtained in \cite{Horowitz:1998ha}
\begin{equation}
ds^{2}=-\frac{r^{2}}{l^{2}}d\tau^{2}+\biggl{(}-\frac{\mu_{s}}{r}+\frac{r^{2}%
}{l^{2}}\biggr{)}^{-1}dr^{2}+\biggl{(}-\frac{\mu_{s}}{r}+\frac{r^{2}}{l^{2}%
}\biggr{)}d\theta^{2}+\frac{r^{2}}{l^{2}}dx_{2}^{2} \label{Schwsol}%
\end{equation}
by using a double analytic continuation $t\rightarrow i\theta$, $x_{1}%
\rightarrow i\tau$ of the planar black hole metric (\ref{bh}). To distinguish
from the black hole solution, we denote by $\mu_{s}$ the mass parameter of the
AdS soliton and, in the Euclidean section ($\tau\rightarrow i\tau_{E}$), the
periodicity is $0\leq\tau_{E}\leq\beta_{s}$. To obtain a regular Lorentzian
solution, the coordinate $r$ is restricted to $r_{s}\leq r$, where
\begin{equation}
-\frac{\mu_{s}}{r_{s}}+\frac{r_{s}^{2}}{l^{2}}=0
\end{equation}
and to avoid the conical singularity in the plane $(r, \theta)$, we impose the
following periodicity for $\theta$:
\begin{equation}
\label{periosol}L_{s}=\frac{4\pi\sqrt{g_{\theta\theta}g_{rr}}}{(g_{\theta
\theta})^{^{\prime}}}\biggr{\vert}_{r=r_{s}}=\frac{4\pi l^{2}}{3r_{s}}%
\end{equation}
The finite on-shell Euclidean action and mass of the AdS soliton can be
obtained in a similar way (but we do not present the details here):
\begin{equation}
I^{E}_{s} = -\frac{LL_{s}\beta_{s}\mu_{s}}{l^{2}} \label{Schwactsol}%
\end{equation}
and the mass can be obtained by using the thermodynamical relations with the
free energy $F=I^{E}_{s}/\beta_{s}=M$ (or from the quasilocal stress tensor)
and the result is
\begin{equation}
M=-\frac{LL_{s}\mu_{s}}{l^{2}}%
\end{equation}
The mass of the AdS soliton corresponds to a Casimir energy associated to the
compact directions of the dual boundary theory, and so it is negative.

With this information it is straightforward to check the existence of first
order phase transitions. To compare the Euclidean solutions, one should impose
the same periodicity conditions, which become in the boundary ($r\rightarrow
\infty$), $\beta_{b}=\beta_{s}$ and $L_{s}=L_{b}$. Let us know compare the
actions (free energies):
\begin{equation}
\Delta I=I_{b}^{E}-I_{s}^{E}=\frac{L}{l^{4}} \biggl{(}\frac{4\pi l^{2}}%
{3}\biggr{)}^{3}L_{b}\beta_{b}[L_{s}^{-3}-\beta_{b}^{-3}]=\frac{L}{l^{4}}
\biggl{(}\frac{4\pi l^{2}}{3}\biggr{)}^{3}L_{b}\beta_{b}\biggl{[}\frac
{1}{L_{s}^{3}}-T^{3}\biggr{]} \label{bhsolfree}%
\end{equation}

The change of sign is an indication of a first order phase transition between
the planar black hole and the AdS soliton. It was shown in \cite{Surya:2001vj}
that the small hot black holes (with respect to $r_{s}$) are unstable and
decay to small hot solitons, but the large cold black holes are stable. Note
that the phase transition is controlled by the dimensionless parameter
$z=TL_{s}$.


\subsection{Hairy AdS soliton in 4-dimensions}

We consider the exact regular hairy black hole solutions with a planar horizon
\cite{Acena:2013jya,Acena:2012mr,Anabalon:2013qua}. The action is
\begin{equation}
I[g_{\mu\nu},\phi]=\int_{\mathcal{M}}{d^{4}x\sqrt{-g}\biggl{[}R-\frac
{(\partial\phi)^{2}}{2}-V(\phi)\biggr{]}}+2\int_{\partial\mathcal{M}}%
{d^{3}xK\sqrt{-h}} \label{action}%
\end{equation}
and we are interested in the following moduli potential:\footnote{For some
particular values of the parameters, it becomes the one of a truncation of
$\omega$-deformed gauged $\mathcal{N}=8$ supergravity \cite{Dall'Agata:2012bb}%
, see \cite{Anabalon:2013eaa, Guarino:2013gsa, Tarrio:2013qga}.}
\begin{align}
V(\phi)  &  =\frac{\Lambda(\nu^{2}-4)}{3\nu^{2}}\biggl{[}\frac{\nu-1}{\nu
+2}e^{-\phi l_{\nu}(\nu+1)}+\frac{\nu+1}{\nu-2}e^{\phi l_{\nu}(\nu-1)}%
+4\frac{\nu^{2}-1}{\nu^{2}-4}e^{-\phi l_{\nu}}\biggr{]}\\
&  +\frac{2\alpha}{\nu^{2}}\biggl{[}\frac{\nu-1}{\nu+2}\sinh{\phi l_{\nu}%
(\nu+1)}-\frac{\nu+1}{\nu-2}\sinh{\phi l_{\nu}(\nu-1)}+4\frac{\nu^{2}-1}%
{\nu^{2}-4}\sinh{\phi l_{\nu}}\biggr{]}\nonumber\\
\nonumber
\end{align}

We focus on the concrete case of $\nu=3$, though hairy AdS solitons for other
values of $\nu$ probably also exist but the analysis is technically more
involved and we do not investigate them in the present work. In this case, the
scalar field potential becomes \newline%
\begin{align}
\label{potential}V(\phi)  &  =\frac{2\Lambda}{27}\biggl{(}5e^{-\phi\sqrt{2}%
}+10e^{\phi\sqrt{2}/2}+16e^{-\phi\sqrt{2}/4}\biggr{)}\\
&  +\frac{4\alpha}{45}\biggl{[}\sinh{\biggl{(}\phi\sqrt{2}\biggr{)}}%
-10\sinh{\biggl{(}\phi\sqrt{2}/2\biggr{)}}+16\sinh{\biggl{(}\phi\sqrt
{2}/4\biggr{)}}\biggr{]}\nonumber\\
\nonumber
\end{align}
The potential has two parts that are controlled by the parameters $\Lambda$
and $\alpha$. Asymptotically, where the scalar field vanishes, just the
parameter $\Lambda$ survives and it relates to the AdS radius as
$\Lambda=-3l^{-2}$.

Using the following metric ansatz
\begin{equation}
ds^{2}=\Omega(x)\left[  -f(x)dt^{2}+\frac{\eta^{2}dx^{2}}{f(x)}+\frac
{dx_{1}^{2}}{l^{2}}+\frac{dx_{2}^{2}}{l^{2}}\right]  \label{Ansatzbh}%
\end{equation}
the equations of motion can be integrated for the conformal factor
\cite{Acena:2013jya, Acena:2012mr, Anabalon:2013qua, Anabalon:2012ta,
Anabalon:2013sra}
\begin{equation}
\Omega(x)=\frac{9x^{2}}{\eta^{2}(x^{3}-1)^{2}} \label{omega}%
\end{equation}
With this choice of the conformal factor, it is straightforward to obtain the
expressions for the scalar field
\begin{equation}
\phi(x)=2\sqrt{2}\ln{x}%
\end{equation}
and metric function%

\begin{equation}
f(x)=\frac{1}{l^{2}}+\alpha\biggl{[}\frac{1}{5}-\frac{x^{2}}{9}%
\biggl{(}1+x^{-3}-\frac{x^{3}}{5}\biggr{)}\biggr{]} \label{f}%
\end{equation}
where $\eta$ is the only integration constant. The parameter $\alpha$ is
positive for $x<1$ and negative otherwise. We shall focus below on the case $x<1$.

The conformal boundary is at $x=1$, where the metric becomes
\begin{equation}
ds^{2}=\frac{R^{2}}{l^{2}}\biggl{[}-dt^{2}+dx_{1}^{2}+dx_{2}^{2}\biggr{]}
\label{Bblack}%
\end{equation}
and we use the following notation for the conformal factor:
\begin{equation}
R^{2}\equiv\frac{1}{\eta^{2}(x-1)^{2}}%
\end{equation}
The geometry where the dual field theory `lives' has the metric
\begin{equation}
ds_{dual}^{2} = \frac{l^{2}}{R^{2}}ds^{2} = \gamma_{ab}dx^{a}dx^{b}%
=-dt^{2}+dx_{1}^{2}+dx_{2}^{2} \label{cftschw}%
\end{equation}
The regularized Euclidean action for these black holes was obtained in
\cite{Anabalon:2015xvl} (see, also, \cite{Anabalon:2016yfg}) (in what follows
we use the same notations as in the previous section for $\beta_{b}$ and
$L_{b}$):
\begin{equation}
I^{E}_{BH}=\beta_{b}\biggr{(}-\frac{\mathcal{A}T}{4G_{N}}+\frac{2LL_{b}}%
{l^{2}}\frac{\alpha}{3\eta^{3}}\biggr{)}=-\frac{LL_{b}\alpha\beta_{b}}%
{3l^{2}\eta^{3}} \label{Ibh}%
\end{equation}
where the area of the horizon and black hole temperature are
\begin{equation}
\mathcal{A}=\frac{LL_{b}\Omega(x_{h})}{l^{2}}, \qquad T=\frac{\alpha}{4\pi
\eta^{3}\Omega}%
\end{equation}
The mass of the hairy black hole is \cite{Anabalon:2015xvl,Anabalon:2014fla}
\begin{equation}
M_{b}= \frac{2LL_{b}\mu_{b}}{l^{2}}, \qquad\mu_{b}=\frac{\alpha}{3\eta^{3}}
\label{massbh}%
\end{equation}
as can be also checked by using the usual thermodynamical relations. Using
this expression of the mass, one can also easily check the first law of thermodynamics.

Let us now construct the hairy AdS soliton. By using again a double analytical
continuation $x_{1}\rightarrow i\tau$ and $t\rightarrow i \theta$ in
(\ref{Ansatzbh}), the metric becomes
\begin{equation}
ds^{2}=\Omega_{s}(x)\left[  -\frac{d\tau^{2}}{l^{2}}+\frac{\lambda^{2}dx^{2}%
}{f(x)}+f(x)d\theta^{2}+\frac{dx_{2} ^{2}}{l^{2}}\right]  . \label{anzatsol}%
\end{equation}
Similarly with the hairy black hole case, the conformal factor (\ref{omega})
is
\begin{equation}
\Omega_{s}(x)=\frac{9x^{2}}{\lambda^{2}(x^{3}-1)^{2}} \label{omegasol}%
\end{equation}
but now we denote the integration constant with $\lambda$ to distinguish it
from the integration constant $\eta$ of the black hole. To get rid of the
conical singularity in the plane $(x,\theta)$, we have to impose the
periodicity:
\begin{equation}
L _{s}=\frac{4\pi\lambda}{f^{^{\prime}}}\biggr{\vert}_{x=x_{s}}=\frac
{4\pi\lambda^{3}\Omega_{s}}{\alpha} \label{periosoliton}%
\end{equation}
where $x_{s}$ is the minimum value of $x$, namely the biggest root of
$f(x_{s})=0$. After imposing the right periodicity on $\theta$ and restricting
the coordinate $x$ so that the metric is Lorentzian, we obtain a well-defined
regular solution.

We use the method of \cite{Anabalon:2015xvl} to compute the regularized
Euclidean action and the result is%

\begin{equation}
I_{soliton}^{E}=-\frac{L\beta_{s}\Omega_{s}(x_{s})}{4l^{2}G_{N}}+\frac
{2LL_{s}\beta_{s}}{l^{2}}\frac{\alpha}{3\lambda^{3}}=-\frac{LL_{s}\beta_{s}%
}{l^{2}}\biggl{(}\frac{\alpha}{3\lambda^{3}}\biggr{)} \label{Isol}%
\end{equation}
from which the mass can be immediately read off:
\begin{equation}
M_{soliton}=-\frac{LL_{s}\mu_{s}}{l^{2}}, \qquad\mu_{s}=\frac{\alpha}%
{3\lambda^{3}} \label{masssoliton}%
\end{equation}
As a check, we have also obtained the quasilocal stress tensor for this case
and then computed the mass, but we do not present the details here.

\subsection{Hairy AdS soliton in 5-dimensions}

Let us now construct an exact hairy AdS soliton solution in five dimensions.
We consider the solutions in \cite{Acena:2012mr}, but we investigate the case
$\nu=5$. In this case, at first sight the potential of \cite{Acena:2012mr} is
not well defined. However, by taking the limit carefully, we obtain that the
theory (potential) and solution are regular. The ansatz metric is
\begin{equation}
ds^{2}=\Omega(x)\left[  -f(x)dt^{2}+\frac{\eta^{2}dx^{2}}{f(x)}+\frac
{dx_{1}^{2}}{l^{2}}+\frac{dx_{2}^{2}}{l^{2}}+\frac{dx_{3}^{2}}{l^{2}}\right]
\label{Ansatz}%
\end{equation}
and, for $\nu=5$, we obtain
\begin{equation}
\Omega(x)=\frac{25x^{4}}{\eta^{2}\left(  x^{5}-1\right)  ^{2}}%
\end{equation}
and
\begin{equation}
f(x)=\frac{1}{l^{2}}+\frac{\alpha}{3^{2}10^{4}}\biggl{(}x^{10}-6x^{5}+30\ln
{x}+3+\frac{2}{x^{5}}\biggr{)}
\end{equation}
The black hole temperature is
\begin{equation}
T=\beta_{b}^{-1}=\left.  \frac{\left\vert \alpha\right\vert }{288\pi\eta
^{4}\left\vert \Omega\right\vert ^{3/2}}\right\vert _{x=x_{h}}%
\end{equation}
where $f(x_{h})=0$. We shall consider the below the case when $\alpha<0$. The
black hole entropy can be also easily computed and we obtain
\begin{equation}
S=\frac{L_{b}L_{2}L_{3}\Omega^{3/2}}{4l^{3}G_{N}}=\frac{\mathcal{A}}{4G_{N}%
},\qquad\mathcal{A}=\frac{L_{b}L_{2}L_{3}}{l^{3}}\Omega^{3/2}%
\end{equation}
To regularize the Euclidean action we choose the following counterterm for the
scalar field:
\begin{equation}
I_{\phi}^{ct}=\int{d^{4}x^{E}\sqrt{h^{E}}}\biggl{(}\frac{\phi^{2}}{2l}%
-\frac{\phi^{3}}{36l}+\frac{7\phi^{4}}{864l}\biggr{)}=\frac{3L_{b}L_{2}L_{3}%
}{l^{3}T}\biggl{[}\frac{6}{\eta^{4}l^{2}(x-1)^{2}}-\frac{8}{\eta^{4}%
l^{2}(x-1)}-\frac{12}{\eta^{4}l^{2}}\biggr{]}+O(x-1)
\end{equation}
The finite action is
\begin{equation}
I_{BH}^{E}=\beta_{b}\biggl{[}-\frac{\mathcal{A}T}{4G_{N}}+\frac{3L_{b}%
L_{2}L_{3}}{l^{3}}\biggl{(}-\frac{\alpha}{288\eta^{4}}\biggr{)}\biggr{]}=\frac
{L_{b}L_{2}L_{3}\beta_{b}}{l^{3}}\biggl{(}\frac{\alpha}{288\eta^{4}}\biggr{)}
\end{equation}
and the mass of the hairy black hole is
\begin{equation}
M_{bh}=-\frac{3L_{b}L_{2}L_{3}\mu_{b}}{l^{3}},\qquad\mu_{b}=\frac{\alpha
}{288\eta^{4}}%
\end{equation}
We again construct the hairy AdS soliton by using a double analytical
continuation $x_{1}\rightarrow i\tau$ and $t\rightarrow i\theta$:
\begin{equation}
ds^{2}=\Omega(x)\left[  -\frac{d\tau^{2}}{l^{2}}+\frac{\eta^{2}dx^{2}}%
{f(x)}+f(x)d\theta^{2}+\frac{dx_{2}^{2}}{l^{2}}+\frac{dx_{3}^{2}}{l^{2}%
}\right]  \label{Ansatz2}%
\end{equation}
The conformal factor for the hairy soliton is
\begin{equation}
\Omega(x)=\frac{25x^{4}}{\lambda^{2}\left(  x^{5}-1\right)  ^{2}}%
\end{equation}
and, to get rid of the conical singularity in the plane $(x,\theta)$, we have
to impose the following periodicity of the angular coordinate:
\begin{equation}
L_{s}=\left\vert \frac{4\pi\lambda}{f^{^{\prime}}}\biggr{\vert}_{x=x_{s}%
}\right\vert =\frac{288\pi\lambda^{4}\Omega^{3/2}}{\left\vert \alpha
\right\vert }%
\end{equation}
We again consider $\alpha<0$, to be consistent with the black hole
case.To complete the analysis, we compute the Euclidean action
\begin{equation}
I_{soliton}^{E}=\beta_{s}\biggl{[}\frac{L_{s}L_{2}L_{3}}{l^{3}}\biggl{(}\frac
{\alpha}{72\lambda^{4}}\biggr{)}+\frac{3L_{s}L_{2}L_{3}}{l^{3}}\biggl{(}-\frac
{\alpha}{288\lambda^{4}}\biggr{)}\biggr{]}=\frac{L_{s}L_{2}L_{3}\beta_{s}%
}{l^{3}}\biggl{(}\frac{\alpha}{288\lambda^{4}}\biggr{)}
\end{equation}
and the mass of the hairy AdS soliton
\begin{equation}
I_{soliton}^{E}\beta_{s}^{-1}=M_{soliton}=\frac{L_{s}L_{2}L_{3}}{l^{3}%
}\biggl{(}\frac{\alpha}{288\lambda^{4}}\biggr{)}
\end{equation}

\section{Implications for phase transitions}

Within AdS/CFT duality, the black holes are interpreted as thermal states in
the dual field theory. We are going to show that there exist first order phase
transitions between the planar hairy black hole and the hairy AdS soliton.

With the results from the previous sections, we are ready to investigate the
existence of phase transitions.\footnote{The case $k=1$, when the horizon
topology is spherical, was studied in \cite{Anabalon:2015ija}.} Let us focus
on $D=4$. Before comparing the actions, we would like to point out that from
the definitions of $x_{s}$ and $x_{h}$ we obtain that they are equal,
$x_{s}=x_{h}$. At first sight, this may be a bit strange because in general it
is expected that they depend on the mass parameters $\lambda$ and $\eta$ for
the soliton and black hole. However, in these unusual coordinates, $x_{s}$ and
$x_{h}$ are defined by (\ref{f}), but the true are of the horizon and
\ `center' of the soliton are determined by the conformal factor in front of
the metric. This conformal factor depends on the mass parameter and we
define:
\begin{equation}
r_{b}^{2}=\frac{\Omega(x_{h},\eta)}{l^{2}},\qquad r_{s}^{2}=\frac{\Omega
(x_{s},\lambda)}{l^{2}}\label{ratio}%
\end{equation}

As before (\ref{bhsolfree}), we have to compare the free energies of solutions
in the same theory and so we have to impose the same periodicity conditions at
the boundary $\beta_{b}=\beta_{s}$ and $L_{s}=L_{b}$. The hairy AdS soliton
has a negative energy (the AdS space in planar coordinates has zero mass) and
it is the ground state of the theory. Hence, the energy of the hairy black
hole should be computed with respect to the ground state and we obtain
\begin{equation}
E=M_{bh}-M_{soliton}=\frac{LL_{b}}{l^{2}}[2\mu_{b}+\mu_{s}]
\end{equation}
with $\mu_{b}$ and $\mu_{s}$ defined in (\ref{massbh}) and(\ref{masssoliton}).

The same periodicity of the Euclidean time implies the same temperature and we
consider the hairy soliton solution as thermal background:
\begin{equation}
\Delta F=\beta_{b}^{-1}(I_{BH}^{E}-I_{soliton}^{E})=\frac{TL\alpha}{3 l^{2}%
}\biggl{(}\frac{L_{s}\beta_{s}}{\lambda^{3}}-\frac{L_{b}\beta_{b}}{\eta^{3}%
}\biggr{)}
\end{equation}
Using the expressions of the black hole temperature $T$ and periodicity
$L_{s}$, we can rewrite the difference of the free energies as
\begin{equation}
\Delta F=\frac{4\pi LL_{s}}{3l^{2}}\biggl{[}\frac{\Omega(\lambda,x_{s})}%
{L_{s}}-T\Omega(\eta,x_{h})\biggr{]} = \frac{4\pi L}{3l^{2}}\Omega
(\lambda,x_{s})\biggl{[}1-\frac{r_{b}^{3}}{r_{s}^{3}}\biggr{]}
\end{equation}
Written in terms of the temperature, there is a drastic change
compared with the no-hair case because the conformal factor appears
explicitly. Clearly, the sign of this expression is controlled by
the ratio $r_{b}/r_{s}$. Interestingly enough, despite the
appearance of the conformal factor, the critical point where $\Delta
F=0$ it is again for the temperature $T_{c}=1/L_{s}$ (that is
because when $\Delta F=0$, $\mu_{b}=\mu_{s}$ and so $\eta=\lambda$).
This is what one expects for a conformal field theory because the
phase transition should depend on the ratio of the scales.

Writing the area of the black hole in terms of $\beta_{b}$ and $\beta_{s}$, we
find that
\begin{equation}
\frac{\mathcal{A}}{Tl^{3}}=\frac{\alpha L}{4\pi l^{5}}\frac{\beta_{b}^{2}%
L_{s}}{\eta^{3}}=\frac{L\mathcal{L}}{l}\biggl{(}\frac{\lambda}{\eta
}\biggr{)}\label{ratiosol}%
\end{equation}
where
\begin{equation}
\mathcal{L}=\frac{16\pi^{2}}{\alpha^{2}l^{4}}\biggl{[}\frac{9x_{h}^{2}}%
{(x_{h}^{3}-1)^{2}}\biggr{]}^{3}%
\end{equation}
However, since $x_{h}$ satisfies $f(x_{h})=0$, it can be computed as a
function of the parameter $\alpha$ of the moduli potential, which implies that
the coefficient $\mathcal{L}\left(  \alpha,l\right)  $ is a function only of
$\alpha$ and $l$. From the definition (\ref{ratio}), one can easily obtain
$r_{b}/r_{s}=\lambda/\eta$ and so (\ref{ratiosol}) can be rewritten in this
useful form:
\begin{equation}
\frac{\mathcal{A}}{Tl^{3}}=\frac{L\,\mathcal{L}\left(  \alpha,l\right)  }%
{l}\frac{r_{b}}{r_{s}}%
\end{equation}

There is an important difference by comparing with the no hair case,
namely the appearance of the function $\mathcal{L}\left(
\alpha,l\right)  $. When $\alpha$ is very small so behave 
$\mathcal{L}$ and, in this case, one can still keep the radius of
the horizon of the same size as $r_{s}$. Therefore, for small
$\alpha$, not only the small hot black holes, but also the large hot
black holes are unstable and decay to hairy AdS solitons. We are
going to comment more on this new feature in `Conclusions' section.
When $\alpha$ parameter is large, the thermodynamical behaviour of
hairy black holes is similar to the one of no-hairy planar black
holes.


\section{Conclusions}

Hawking and Page have shown that there exists a phase transition between
spherical AdS (Schwarzschild) black hole and global ($k=1$) AdS spacetime. As
is well known, the phase transition, both on the gravity side and on the gauge
theory side, is sensitive to the topology of the AdS foliation. For AdS black
holes with planar horizon geometry, there exists no Hawking-Page transition
with respect to AdS spacetime. In other words, the planar black hole phase is
always dominant for any non-zero temperature.

Interestingly, it was shown that when one (or more directions) are compact
there exist also Hawking-Page phase transitions between the planar black holes
and the AdS soliton, which is obtained by a double analytic continuation from
the black hole. We have obtained a similar behaviour for the hairy black
holes, but now the ground state corresponds to a hairy soliton. One important
difference with the no hair case is that the phase transition is also
controlled by the parameter $\alpha$ in the scalar potential. Once $\alpha$ is
fixed, the theory is fixed, but for very small $\alpha$ the theory contains
hot black holes (small or large) that are unstable and decay to hairy AdS
solitons. This drastic change is related to the fact that when $\alpha$
vanishes, the hairy black hole solutions become naked singularities. The self
interaction of the scalar fied is very weak and so a large temperature can
destabilize the system regardless of the size of the black hole.

As a future direction, it will be interesting to understand the physics of
this instability in the dual field theory. It will also be interesting to
investigate the general phase diagram for an arbitrarily parameter $\nu$ in
the moduli potential and the embedding in supergravity \cite{noi}. When the effective cosmological constant vanishes, one can also
obtain hairy black holes in flat space (stationary hairy black holes were also
obtained, but only numerically). The thermodynamics and phase diagram of
asymptotically flat hairy black holes
\cite{Anabalon:2013qua,Herdeiro:2015waa,Herdeiro:2015gia,Herdeiro:2014goa} can
be also studied with a similar counterterm method
\cite{Astefanesei:2005ad,Mann:2005yr,Astefanesei:2009wi}.

\section*{Acknowledgments} Research of AA is supported in part by
FONDECYT Grants 1141073 and 1161418 and Newton-Picarte Grants
DPI20140053 and DPI20140115. The work of DA is supported by the
FONDECYT Grant 1161418 and Newton-Picarte Grant DPI20140115.


\newpage

\begin{thebibliography}{99}                                                                                               %


\bibitem {Maldacena:1997re}J.~M.~Maldacena, ``The Large N limit of
superconformal field theories and supergravity,''
Int.\ J.\ Theor.\ Phys.\ \textbf{38}, 1113 (1999)
[Adv.\ Theor.\ Math.\ Phys.\ \textbf{2}, 231 (1998)]
doi:10.1023/A:1026654312961 [hep-th/9711200].

\bibitem {Witten:1998zw}E.~Witten, ``Anti-de Sitter space, thermal phase
transition, and confinement in gauge theories,''
Adv.\ Theor.\ Math.\ Phys.\ \textbf{2}, 505 (1998) [hep-th/9803131].


\bibitem {Hertog:2004ns}T.~Hertog and G.~T.~Horowitz, ``Designer gravity and
field theory effective potentials,'' Phys.\ Rev.\ Lett.\ \textbf{94}, 221301
(2005) doi:10.1103/PhysRevLett.94.221301 [hep-th/0412169].

\bibitem {Hertog:2004rz}T.~Hertog and G.~T.~Horowitz, ``Towards a big crunch
dual,'' JHEP \textbf{0407}, 073 (2004) doi:10.1088/1126-6708/2004/07/073 [hep-th/0406134].

\bibitem {Anabalon:2015vda}A.~Anabalon, D.~Astefanesei and J.~Oliva, ``Hairy
Black Hole Stability in AdS, Quantum Mechanics on the Half-Line and
Holography,'' JHEP \textbf{1510}, 068 (2015) doi:10.1007/JHEP10(2015)068
[arXiv:1507.05520 [hep-th]].

\bibitem {Hawking}S.~W.~Hawking and D.~Page, ``Thermodynamics Of Black Holes
In Anti-de Sitter Space,'' Commun.~Math.~Phys.\ \textbf{87} (1983) 577.

\bibitem {Horowitz:1998ha}G.~T.~Horowitz and R.~C.~Myers, ``The AdS / CFT
correspondence and a new positive energy conjecture for general relativity,''
Phys.\ Rev.\ D \textbf{59}, 026005 (1998) doi:10.1103/PhysRevD.59.026005 [hep-th/9808079].

\bibitem {Brihaye:2013tra}Y.~Brihaye, B.~Hartmann and S.~Tojiev, ``AdS
solitons with conformal scalar hair,'' Phys.\ Rev.\ D \textbf{88}, 104006
(2013) doi:10.1103/PhysRevD.88.104006 [arXiv:1307.6241 [gr-qc]].

\bibitem {Ogawa:2011fw}N.~Ogawa and T.~Takayanagi, ``Higher Derivative
Corrections to Holographic Entanglement Entropy for AdS Solitons,'' JHEP
\textbf{1110}, 147 (2011) doi:10.1007/JHEP10(2011)147 [arXiv:1107.4363 [hep-th]].

\bibitem {Shi:2016bxz}H.~q.~Shi and D.~f.~Zeng, ``Geodesic Motions in AdS
Soliton Background Space-time,'' arXiv:1603.08624 [hep-th].

\bibitem {Kleihaus:2013tba}B.~Kleihaus, J.~Kunz, E.~Radu and B.~Subagyo,
``Axially symmetric static scalar solitons and black holes with scalar hair,''
Phys.\ Lett.\ B \textbf{725}, 489 (2013) doi:10.1016/j.physletb.2013.07.051

\bibitem {Brihaye:2012ww}Y.~Brihaye, B.~Hartmann and S.~Tojiev, ``Formation of
scalar hair on Gauss-Bonnet solitons and black holes,'' Phys.\ Rev.\ D
\textbf{87}, no. 2, 024040 (2013) doi:10.1103/PhysRevD.87.024040
[arXiv:1210.2268 [gr-qc]].

\bibitem {Smolic:2015txa}I.~Smoli\"{A}\ddag, ``Symmetry inheritance of scalar
fields,'' Class.\ Quant.\ Grav.\ \textbf{32}, no. 14, 145010 (2015)
doi:10.1088/0264-9381/32/14/145010 [arXiv:1501.04967 [gr-qc]].

\bibitem {Cadoni:2011yj}M.~Cadoni, S.~Mignemi and M.~Serra, ``Black brane
solutions and their solitonic extremal limit in Einstein-scalar gravity,''
Phys.\ Rev.\ D \textbf{85}, 086001 (2012) doi:10.1103/PhysRevD.85.086001
[arXiv:1111.6581 [hep-th]].

\bibitem {Acena:2013jya}A.~Ace\~{A}$\pm$a, A.~Anabal\~{A}%
${{}^3}$%
n, D.~Astefanesei and R.~Mann, ``Hairy planar black holes in higher
dimensions,'' JHEP \textbf{1401}, 153 (2014) doi:10.1007/JHEP01(2014)153
[arXiv:1311.6065 [hep-th]].

\bibitem {Acena:2012mr}A.~Acena, A.~Anabalon and D.~Astefanesei, ``Exact hairy
black brane solutions in $AdS_{5}$ and holographic RG flows,'' Phys.\ Rev.\ D
\textbf{87}, no. 12, 124033 (2013) doi:10.1103/PhysRevD.87.124033
[arXiv:1211.6126 [hep-th]].

\bibitem {Feng:2013tza}X.~H.~Feng, H.~Lu and Q.~Wen, ``Scalar Hairy Black
Holes in General Dimensions,'' Phys.\ Rev.\ D \textbf{89}, no. 4, 044014
(2014) doi:10.1103/PhysRevD.89.044014 [arXiv:1312.5374 [hep-th]].

\bibitem {Wen:2015xea}Q.~Wen, ``Strategy to Construct Exact Solutions in
Einstein-Scalar Gravities,'' Phys.\ Rev.\ D \textbf{92}, no. 10, 104002 (2015)
doi:10.1103/PhysRevD.92.104002 [arXiv:1501.02829 [hep-th]].

\bibitem {Lu:2013ura}H.~L\~{A}%
${\frac14}$%
, Y.~Pang and C.~N.~Pope, ``AdS Dyonic Black Hole and its Thermodynamics,''
JHEP \textbf{1311}, 033 (2013) doi:10.1007/JHEP11(2013)033 [arXiv:1307.6243 [hep-th]].

\bibitem {Fan:2015tua}Z.~Y.~Fan and H.~Lu, ``Static and Dynamic Hairy Planar
Black Holes,'' Phys.\ Rev.\ D \textbf{92}, no. 6, 064008 (2015)
doi:10.1103/PhysRevD.92.064008 [arXiv:1505.03557 [hep-th]].

\bibitem {Fan:2015ykb}Z.~Y.~Fan and B.~Chen, ``Exact formation of hairy planar
black holes,'' Phys.\ Rev.\ D \textbf{93}, no. 8, 084013 (2016)
doi:10.1103/PhysRevD.93.084013 [arXiv:1512.09145 [hep-th]].

\bibitem {Woolgar:2016axs}E.~Woolgar, ``The rigid Horowitz-Myers conjecture,''
arXiv:1602.06197 [math.DG].

\bibitem {Galloway:2001uv}G.~J.~Galloway, S.~Surya and E.~Woolgar, ``A
Uniqueness theorem for the AdS soliton,'' Phys.\ Rev.\ Lett.\ \textbf{88},
101102 (2002) doi:10.1103/PhysRevLett.88.101102 [hep-th/0108170].

\bibitem {Galloway:2002ai}G.~J.~Galloway, S.~Surya and E.~Woolgar, ``On the
geometry and mass of static, asymptotically AdS space-times, and the
uniqueness of the AdS soliton,'' Commun.\ Math.\ Phys.\ \textbf{241}, 1 (2003)
doi:10.1007/s00220-003-0912-7 [hep-th/0204081].

\bibitem {Surya:2001vj}S.~Surya, K.~Schleich and D.~M.~Witt, ``Phase
transitions for flat AdS black holes,'' Phys.\ Rev.\ Lett.\ \textbf{86}, 5231
(2001) doi:10.1103/PhysRevLett.86.5231 [hep-th/0101134].

\bibitem {Balasubramanian:1999re}V.~Balasubramanian and P.~Kraus, ``A Stress
tensor for Anti-de Sitter gravity,'' Commun.\ Math.\ Phys.\ \textbf{208}, 413
(1999) doi:10.1007/s002200050764 [hep-th/9902121].

\bibitem {Anabalon:2015xvl}A.~Anabalon, D.~Astefanesei, D.~Choque and
C.~Martinez, ``Trace Anomaly and Counterterms in Designer Gravity,'' JHEP
\textbf{1603}, 117 (2016) doi:10.1007/JHEP03(2016)117 [arXiv:1511.08759 [hep-th]].

\bibitem {Anabalon:2013qua}A.~Anabalon, D.~Astefanesei and R.~Mann, ``Exact
asymptotically flat charged hairy black holes with a dilaton potential,'' JHEP
\textbf{1310}, 184 (2013) doi:10.1007/JHEP10(2013)184 [arXiv:1308.1693 [hep-th]].

\bibitem {Dall'Agata:2012bb}G.~Dall'Agata, G.~Inverso and M.~Trigiante,
``Evidence for a family of SO(8) gauged supergravity theories,''
Phys.\ Rev.\ Lett.\ \textbf{109}, 201301 (2012)
doi:10.1103/PhysRevLett.109.201301 [arXiv:1209.0760 [hep-th]].

\bibitem {Anabalon:2013eaa}A.~Anabalon and D.~Astefanesei, ``Black holes in
$\omega$-defomed gauged $N=8$ supergravity,'' Phys.\ Lett.\ B \textbf{732},
137 (2014) doi:10.1016/j.physletb.2014.03.035 [arXiv:1311.7459 [hep-th]].

\bibitem {Guarino:2013gsa}A.~Guarino, ``On new maximal supergravity and its
BPS domain-walls,'' JHEP \textbf{1402}, 026 (2014) doi:10.1007/JHEP02(2014)026
[arXiv:1311.0785 [hep-th]].

\bibitem {Tarrio:2013qga}J.~Tarr\~{A}-o and O.~Varela, ``Electric/magnetic
duality and RG flows in AdS$_{4}$/CFT$_{3}$,'' JHEP \textbf{1401}, 071 (2014)
Addendum: [JHEP \textbf{1512}, 068 (2015)] doi:10.1007/JHEP01(2014)071,
10.1007/JHEP12(2015)068 [arXiv:1311.2933 [hep-th]].

\bibitem {Anabalon:2012ta}A.~Anabalon, ``Exact Black Holes and Universality in
the Backreaction of non-linear Sigma Models with a potential in (A)dS4,'' JHEP
\textbf{1206}, 127 (2012) doi:10.1007/JHEP06(2012)127 [arXiv:1204.2720 [hep-th]].

\bibitem {Anabalon:2013sra}A.~Anabal\~{A}%
${{}^3}$%
n and D.~Astefanesei, ``On attractor mechanism of $AdS_{4}$ black holes,''
Phys.\ Lett.\ B \textbf{727}, 568 (2013) doi:10.1016/j.physletb.2013.11.013
[arXiv:1309.5863 [hep-th]].

\bibitem {Anabalon:2016yfg}A.~Anabalon, D.~Astefanesei and R.~Mann,
``Holographic equation of state in fluid/gravity duality,'' arXiv:1604.05595 [hep-th].

\bibitem {Anabalon:2014fla}A.~Anabalon, D.~Astefanesei and C.~Martinez, ``Mass
of asymptotically anti\^{a} de Sitter hairy spacetimes,'' Phys.\ Rev.\ D
\textbf{91}, no. 4, 041501 (2015) doi:10.1103/PhysRevD.91.041501
[arXiv:1407.3296 [hep-th]].

\bibitem {Anabalon:2015ija}A.~Anabalon, D.~Astefanesei and D.~Choque, ``On the
thermodynamics of hairy black holes,'' Phys.\ Lett.\ B \textbf{743}, 154
(2015) doi:10.1016/j.physletb.2015.02.024 [arXiv:1501.04252 [hep-th]].

\bibitem {noi}``work in progress''.

\bibitem {Herdeiro:2015waa}C.~A.~R.~Herdeiro and E.~Radu, ``Asymptotically
flat black holes with scalar hair: a review,'' Int.\ J.\ Mod.\ Phys.\ D
\textbf{24}, no. 09, 1542014 (2015) doi:10.1142/S0218271815420146
[arXiv:1504.08209 [gr-qc]].

\bibitem {Herdeiro:2015gia}C.~Herdeiro and E.~Radu, ``Construction and
physical properties of Kerr black holes with scalar hair,''
Class.\ Quant.\ Grav.\ \textbf{32}, no. 14, 144001 (2015)
doi:10.1088/0264-9381/32/14/144001 [arXiv:1501.04319 [gr-qc]].

\bibitem {Herdeiro:2014goa}C.~A.~R.~Herdeiro and E.~Radu, ``Kerr black holes
with scalar hair,'' Phys.\ Rev.\ Lett.\ \textbf{112}, 221101 (2014)
doi:10.1103/PhysRevLett.112.221101 [arXiv:1403.2757 [gr-qc]].

\bibitem {Astefanesei:2005ad}D.~Astefanesei and E.~Radu, ``Quasilocal
formalism and black ring thermodynamics,'' Phys.\ Rev.\ D \textbf{73}, 044014
(2006) doi:10.1103/PhysRevD.73.044014 [hep-th/0509144].

\bibitem {Mann:2005yr}R.~B.~Mann and D.~Marolf, ``Holographic renormalization
of asymptotically flat spacetimes,'' Class.\ Quant.\ Grav.\ \textbf{23}, 2927
(2006) doi:10.1088/0264-9381/23/9/010 [hep-th/0511096].

\bibitem {Astefanesei:2009wi}D.~Astefanesei, R.~B.~Mann, M.~J.~Rodriguez and
C.~Stelea, ``Quasilocal formalism and thermodynamics of asymptotically flat
black objects,'' Class.\ Quant.\ Grav.\ \textbf{27}, 165004 (2010)
doi:10.1088/0264-9381/27/16/165004 [arXiv:0909.3852 [hep-th]].



\end{thebibliography}
\end{document}